\documentclass{article}
\def\bea{\begin{eqnarray}}
\def\eea{\end{eqnarray}}
\def\nn{\nonumber}

\def\5{\hspace*{5mm}}

\def\bx{{{{\bf x}}}}

\def\rd{{\rm d}}

\def\p{\partial}
\def\a{\alpha}
\def\b{\beta}

\def\d{\delta}

\def\half{\frac12}

\def\ri{{\rm i}}
\def\re{{\rm e}}

\def\ie{{\it i.e.}}
\def\bz{{\bar z}}

\def\ei{\epsilon^{ik}}

\def\sma{{\sum_{\a=1}^N}}
\def\smb{\sum_{{\b=1},\hspace{0,02cm}{\b\not=\a}}^{N}}
\def\A'{A^\star}
\def\B'{B^\star}

\begin{document}

\begin{center}
{\large\bf Geometric Transformations and NCCS Theory\\
in the Lowest Landau Level}

\vspace*{5mm}

M. Eliashvili and G. Tsitsishvili\\

\vspace*{5mm}

{\it Department of Theoretical Physics,\\
A. Razmadze Mathematical Institute,\\ Tbilisi 380093 Georgia\\
simi@rmi.acnet.ge}

\end{center}

\begin{abstract}
Chern-Simons type gauge field is generated by the means of the
singular area preserving transformations in the lowest Landau
level of electrons forming fractional quantum Hall state. Dynamics
is governed by the system of constraints which correspond to the
Gauss law in the non-commutative Chern-Simons gauge theory and to
the lowest Landau level condition in the picture of composite
fermions. Physically reasonable solution to this constraints
corresponds to the Laughlin state. It is argued that the model
leads to the non-commutative Chern-Simons theory of the QHE and
composite fermions.
\end{abstract}

\vspace*{5mm}
\noindent
{\bf 1. Introduction}
\vspace*{3mm}

One of the intriguing features of the quantum Hall effect (QHE)
(for a comprehensive introduction see Ref. 1) is that it is a
simplest physical realization of the non-commutative spatial
geometry (see \textit{e.g.} Ref. 2). Due to the intense orthogonal
magnetic field $\mathbf{B}=(0,0,B_{\bot })$, electrons are confined
to the lowest Landau level (LLL) and their position coordinates do
not commute:
\bea
\lbrack\hat{x},\hat{y}]=\frac{\ri}{B_{\bot}}\equiv-\ri\theta.
\eea

This fact had stimulated a considerable number of papers, in which
quantum Hall effect is examined from the point of view of the
non-commutative quantum field theory \cite{susskind}$^-$\cite{guralnik}.

In the Ref. 3 it was shown, that the Laughlin's theory \cite{laughlin}
of fractional quantum Hall effect (FQHE) for the odd inverse filling
factors $\nu^{-1}=2p+1$ can be presented as a non-commutative
Chern-Simons (NCCS) gauge theory. This assertion is formulated in
terms of the fluid mechanics and is based on the use of hydrodynamical
variables introduced in Ref. 12.

It would be interesting to substantiate the above assertion using
mechanical variables -- electron coordinates and momenta. This could
make more transparent a transition to the traditional quantum
mechanical description of the electron system. Another interesting
question is how fits spatial non-commutativity with a composite
fermion (CF) picture \cite{jain}, according which the fractional
quantum Hall states are formed by quasiparticles experiencing the
reduced magnetic field ${B^\star_{\bot }}=\nu{B_{\bot }}$. Hence one
has to take into account that in the CF picture the non-commutativity
parameter will be $\theta^\star={\nu}^{-1}\theta$.

In the field theory the composite particle  scenario can be introduced
by the means of the Chern-Simons (CS) gauge theory \cite{lopez,zhang}.
Here the central role is played by the Gauss's law -- constraint
binding the CS magnetic field to the electron density and providing a
flux attachment mechanism. In Ref. 3 it was obtained the Gauss's
law for the CS theory based on the group of area preserving
diffeomorphysms \cite{manvelyan} (APD) with a subsequent interpretation
of it as a first order truncation of the corresponding non-commutative
theory.

In the present paper we discuss these questions considering electrons
as charged particles with zero kinetic momenta, {\it i.e.} in the LLL.
In Ref. 17 it was argued that the standard CS approach and CF picture
can be developed starting with some area preserving singular geometric
transformations and considering them in the context of QHE. Below we
will show, that in reality the area invariance condition leads to the
Gauss's law for CS theory with APD group. We also propose a simple
modification of the area transformation rule, which will permit to
write down the Gauss's law directly for the full non-linear NCCS theory.
Area transformations induce corresponding changes in the $N$-particle
Lagrangian and can be interpreted as a transition to the CF picture in
the framework of the NCCS theory.

The lay-out of this paper is as follows. In Section 2 we introduce
transformations of the electron Lagrangian and relate them to the
area preserving transformations. In remaining sections 3 and 4 the
corresponding NCCS theory is considered and the simple solution for
the Laughlin state is determined.

\textbf{Notations:}. In the $x-y$ plane together with the Cartesian
coordinates $x^i=x_i$ ($i=1,2$) we use the complex ones: $z=x+iy$,
$\bar z=x-iy$. In the natural units $c=\hbar=1$ electrons have
a charge $e=-1$, mass $m$ and move in the area $\Omega$ in the
homogeneous magnetic field pointing down the $\hat{\mathbf{z}}$
axis: $B_{\bot}=\partial_x A_y-\partial_y A_x=-B<0$. For 2+1
space-time we use coordinates $x^\mu$ $(x^0=t,x^i)$ and metric
tensor $\eta_{\mu\nu}=diag(-1,+1,+1)$.

\vspace*{5mm}
\noindent
{\bf 2. Electrons in LLL and Geometric Transformations}
\vspace*{3mm}

Consider electrons moving in the $x$ -- $y$ plane in the presence of the
intense orthogonal magnetic field. In the case of quantum Hall states
with the filling factor
\bea
\nu =\frac{2\pi N}{B\Omega}=\frac{1}{2p+1} \label{eq:ff1}
\eea
one deals with the incompressible quantum fluid formed by LLL
electrons occupying the area $\Omega$. In what follows we mainly
consider a finite system of electrons of a limited spatial extent.

In the symmetric gauge where $A_i=\frac12 B\epsilon_{ik}x^k$, the
$N$-particle LLL wave function must satisfy equations
\begin{eqnarray}
\hat\pi_\bz^\a\Psi({{\bf x}}_1,...,{{\bf x}}_N)\equiv
\{\hat p_\bz^\a+A_\bz(\mathbf{r}_\a)\}
\Psi({{\bf x}}_1,...,{{\bf x}}_N)=0. \label{eq:LLL}
\end{eqnarray}
Here the locus of particles is described by their position vectors
${{{\bf x}}}_\a$ ($\a=1,2,...N$). Solution to the Eq.(\ref{eq:LLL})
is of the form
\begin{eqnarray}
\Psi({{\bf x}}_1,...,{{\bf x}}_N)=F(z_1,...,z_N)
\re^{-\frac{B}{4}{\sum_\a}|z_\a|^2}  \label{eq:WF}
\end{eqnarray}
and the dynamical information on the system is encapsuled in the
holomorphic function $F(z_1,...,z_N)$.

The equation (\ref{eq:LLL}) can be viewed as a condition
imposed on the physical states by the constraint dynamics
(in the Dirac's \cite{dirac} sense), and at the classical
level these constraints lead to the vanishing  kinetic momenta:
\begin{eqnarray}
\pi^\a_i\approx 0.  \label{eq:dc}
\end{eqnarray}
Constraints are of the second class with the immediate consequence
that the electron coordinates do not commute:
\bea
[\hat x_\a,\hat
y_\b]_-=\ri\{x_\a,y_\b\}_{Dirac}=-\ri\theta\d_{\a\b}.
\eea
Interesting to note that the Dirac bracket for the particle
densities is also non-zero \cite{guralnik}
\bea
\{\rho({{{\bf x}}}),\rho({{{\bf x}}}')\}_{Dirac}
=-\theta\epsilon_{ik}\frac{\p}{\p x^i_\a}\rho({{{\bf x}}})
\frac{\p}{\p x^k_\a}\rho({{{\bf x}}}').
\eea

Constraints $\pi^\a_i\approx 0$ are provided by the singular
Lagrangian \cite{dunne}
\begin{eqnarray}
L_N=-\sum_{\alpha =1}^{N}\dot{x}_{\alpha}^{i}A_i({{\bf x}}_\a)
=-\frac{B}{2}\sma \dot x^i_\a\epsilon_{ik}x^k_\a.
\label{eq:lagN}
\end{eqnarray}

In the limit of the strong magnetic field one can neglect the
kinetic term, {\it i.e} formally put $m=0$. In (\ref{eq:lagN}) we
have not included different interaction terms (electron-electron,
electron-background {\it etc}.) or confinement forces and concentrate
on the terms which are of first order in time derivatives.

The corresponding canonical Hamiltonian vanishes and the quantum
dynamics is completely governed by constraints (\ref {eq:dc}). At
the same time one cannot impose operator constraints
\begin{eqnarray}
\hat\pi^\a_i|\psi_L\rangle=0\hspace*{10mm}\langle\psi_L|\hat\pi^\a_i=0,
\nonumber
\end{eqnarray}
because they do not commute among themselves
\begin{eqnarray}
0=\langle\psi_L|[\hat\pi^\a_1,\hat\pi^\a_2]
|\psi_L\rangle=\ri B\ne0.
\nonumber
\end{eqnarray}
Instead, constraints can vanish only "weakly", {\textit{i.e.}}
\begin{eqnarray}
\hat\pi^\a_{\bar z}|\psi_L\rangle =0,
\hspace*{10mm}\langle\psi_L|\hat\pi^\a_z=0.
\nonumber
\end{eqnarray}
This is consistent with classical Eqs. (\ref{eq:dc}), as well as
the corresponding quantum averages vanish
\begin{eqnarray}
\langle\Psi_L|\hat\pi^\a_i|\Psi_L\rangle=0. \nonumber
\end{eqnarray}

Remark now, that the formal substitution
$B\rightarrow B^{\star}\equiv \nu B$, $\Omega\rightarrow\Omega$
amounts to the change of the filling factor
$\nu\rightarrow\nu^{\star }=1$. Referred to above substitution
can be expressed in terms of the transformations of particle
coordinates and velocities
\begin{eqnarray}
x^i_\a\rightarrow\sqrt{\nu}\{x_\a^i-\theta^\star\epsilon^{ik}
a_k^\a({\bf x}_1,...,{\bf x}_N)\}  \label{eq:tr1}
\end{eqnarray}
\bea
\dot x^i_\a\rightarrow\sqrt{\nu}\{\dot x_\a^i-\theta^\star
\epsilon^{ik}\dot a^\a_k({\bf x}_1,...,{\bf x}_N)\} \label{eq:tr2}
\eea
where $a_i^\a$ and $\dot a_i^\a$ are yet unspecified functions of
particle coordinates.

Under (\ref{eq:tr1}) and (\ref{eq:tr2}) Lagrangian (\ref{eq:lagN})
is transformed to
\begin{eqnarray}
L^{\star}_N=-{\sum_{\alpha=1}^N}\dot x^i_\a\big\{A^\star_i
({{\bf x}}_\a)+ a^\a_i({{\bf x}}_\a)\big\}+\frac{\theta^\star}{2}
\sma\epsilon^{ik}a_i^\a\dot a_k^\a. \label{eq:lstar}
\end{eqnarray}

In order to specify functions $a_i^\a$ and $\dot a_i^\a$ we appeal
to the physical properties of the Hall fluid. First of all, the system
of electrons moving in the strong magnetic field forms a special
kind of incompressible quantum fluid and incompressibility can be
formulated in terms of the area preserving diffeomorphisms, which
is the symmetry of non-interacting electrons in the magnetic
field \cite{cappelli,karabali}. Secondly, according to the CF
picture \cite{jain} electrons in the quantum Hall state are replaced
by the so called composite particles -- fermions (bosons) carrying
even (odd) number of elementary magnetic flux quantum. Composite
particles experience the effective magnetic field $B^\star=\nu B$ and
they fill up their own lowest Landau level.

In the present paper we argue that the emergence of the Chern-Simons
gauge field can be understood considering the area preserving
transformations, {\ie} CS field has a geometric origin. In order
to clarify this point let us turn to the incompressibility and its
geometric manifestation. Particles are restricted to move in the area:
\begin{eqnarray}
\Omega=\int_D d^2x=(2p+1)\frac{2\pi}{B}N. \label{eq:Area}
\end{eqnarray}
Consider the map
\bea
x^i\rightarrow {x'}^i=F^i(\bx), \label{eq:map}
\eea
which induces the change of the area (\ref{eq:Area}):
\bea
\Omega\rightarrow\Omega'=\int_D d^2 x \mathcal{J}
\Big(\frac{\partial F^i}{\partial x^k}\Big), \label{eq:jacob}
\eea
where
\bea
\mathcal{J}\Big(\frac{\partial F^i}{\partial x^k}\Big)=
\half\epsilon_{ik}\epsilon^{mn}\p_m F^i\p_n F^k
\eea
is a Jacobian of the transformation (\ref{eq:map})

Introduce a deformed multiplication of two functions
\bea
(f\circ g)_\tau=fg-\ri\frac{\tau}{2}\epsilon^{mn}
\p_m f\p_n g \label{eq:circ}
\eea
where $\tau$ is some parameter. Then the area transformation
can be presented in the following form
\bea
\Omega=\frac{\ri}{\theta}\int_D\rd^2x \epsilon_{ik}
(x^i\circ x^k)_\theta\rightarrow\Omega'=\frac{\ri}{\theta^\star}
\int_D\rd^2x \epsilon_{ik}(F^i\circ F^k)_{\theta^\star}.
\eea
Here we admit, that the map (\ref{eq:map}) is accompanied by the
parameter change $\theta\rightarrow \theta^\star$.

The area preservation condition looks as follows
\bea
\Omega_\theta\equiv \Omega\rightarrow\Omega'\equiv
\Omega_{\theta^\star}. \label{inv1}
\eea

Let
\begin{eqnarray}
F^i({\bf x})= s(i)\{x^i-\theta^\star\epsilon^{ik} a_k({\bf x})\}
\label{eq:tr3}
\end{eqnarray}
where $a_k(\bx)$ is a deformation field and $s(i)$'s are some constants.

Taking into account the definition of the filling factor
(\ref{eq:ff1}) one gets
\bea
\frac{1-s}{s}\frac{2\pi N}{\nu B}=-\theta^\star\int_D\rd^2x
\ei D_ia_k. \label{eq:area1}
\eea
Here $s=s(1)s(2)$ and we use the notation
\bea
D_ia_k=\partial_ia_k-\ri(a_i\circ a_k)_{\theta^\star}.
\label{eq:curl}
\eea
Setting $s=\nu=(1+2p)^{-1}$ we end up with the equation
\begin{eqnarray}
\int_{D}{\rd}^2x\epsilon^{ik}D_ia_k=-4\pi pN. \label{eq:deform}
\end{eqnarray}

Introduce the local density $\rho({{\bf x}},t)$, satisfying the
condition
\begin{eqnarray}
N=\int_D{\rd}^2x\rho({{\bf x}},t). \label{eq:density}
\end{eqnarray}
In (\ref{eq:density}) we admit that the density can be
time-dependent.

Now the integral relation (\ref{eq:deform}) can be written
in the form of local constraint imposed on the deformation
field $a_i(\bx)$
\begin{eqnarray}
\rho({{{\bf x}}},t)+\kappa\ei D_ia_k=0,
\hspace*{10mm} 1/\kappa =4\pi p. \label{eq:gauss}
\end{eqnarray}
The last relation resembles the Gauss law known in the Chern-Simons
gauge field theory. The main difference is that in (\ref{eq:gauss})
we use the "covariant curl" (\ref{eq:curl}), differing from the usual
one by the non-linear term
$\epsilon ^{ik} (a_{i}\circ a_{k})_{\theta^\star}$.

Denote the solution of Eq. (\ref{eq:gauss}) by
$a_i({\bf x}|{{\bf x}}_1,...,{{\bf x}}_N)$ and formally define
\bea
a^\a_k=a^\a_k(\bx|\bx_1,...,\bx_N)|_{\bx=\bx_\a}\equiv
a_k(\bx_\a)\label{eq:a}
\eea
and
\bea
\dot a^\a_k=\frac{1}{2p}\frac{\p}{\p t}a_k(\bx|\bx_1,...,\bx_N)
|_{\bx=\bx_\a}\equiv\frac{1}{2p}\dot a_k(\bx_\a). \label{eq:da}
\eea

Substitution of (\ref{eq:a}) and (\ref{eq:da}) into Lagrangian
(\ref{eq:lstar}) yields
\bea
L_N^\star=-{\sum_{\alpha=1}^N} \dot x^i_\a
\big\{A^\star_i({{\bf x}}_\a)+ a_i({{\bf x}}_\a)\big\}
+\frac{\theta^\star}{2}\frac{1}{2p}\frac{1}{\Delta}\sma\Delta\ei
a_i({{{{\bf x}}}}_\a)\dot a_k({{{{\bf x}}}}_a)\label{eq:lstar1}
\eea
where $\Delta=2\pi/\B'$.

If we suppose that the sum in (\ref{eq:lstar1}) may be replaced
by the integral, {\ie}
\bea
\sma \ei \Delta \ei a_i({\bf x}_\a)\dot a_k({\bf x}_a)
\rightarrow\int_D{{\rd}^2x}\ei a_i({\bf x},t)\p_t a_k({\bf x},t)
\label{eq:integral}
\eea
we arrive at the Lagrangian
\bea
L^\star=\int_D{\rd}^2x\bigg\{-j^i({\bf x},t)[A^\star_i({\bf x})
+a_i({\bf x},t)]+\frac{\kappa}{2}\ei a_i({\bf x},t)\p_t
a_k({\bf x},t)\bigg\}. \label{eq:lstar2}
\eea
Validity of the substitution (\ref{eq:integral}) can be corroborated
by the fact that $\Delta$ is an elementary area occupied by the
CF in the magnetic field $\B'$.

In the Lagrangian (\ref{eq:lstar2})
\bea
j^i({{{\bf x}}},t)=\sma\dot x^i_\a(t)\d({{{{\bf x}}}-{{{\bf x}}}_\a})
\eea
is a 2-current and the
field $a_i({\bf x},t)$ is subject to the Gauss law (\ref{eq:gauss})
\begin{eqnarray}
\Phi\equiv\rho({\bf x},t)+\kappa\ei D_ia_k=0\label{eq:ngl}
\end{eqnarray}

Remark that the constraint (\ref{eq:ngl}) is invariant under the
infinitesimal gauge transformations
\bea
\d a_i({\bf x},t)=\p_i\lambda+\ri\{(\lambda\circ a_i)_{\theta^\star}
-(a_i\circ\lambda)_{\theta^\star}\}\label{eq:gtr}
\eea
\bea
\d\rho({\bf x},t)=\ri\{(\lambda\circ\rho)_{\theta^\star}-
(\rho\circ\lambda)_{\theta^\star}\}
\eea

The Gauss law $\Phi=0$ can be taken into account considering
Lagrangian
\bea
L&=&\int_D{\rd}^2x\bigg\{-j^i({\bf x},t)(A^\star_i({\bf x})+a_i({\bf x},t))
+\frac{\kappa}{2}\ei a_i({\bf x},t)\p_t a_k({\bx},t)\nn\\
&-&a_0({\bf x},t)\Phi({\bf x},t)\bigg\}.
\eea
Variation with respect to the Lagrange multiplier field $a_0$ imposes
the constraint (\ref{eq:ngl}).

Assuming that on the boundary $a_0({{{\bf x}}})|_{\p D}=0$ the
corresponding Lagrangian density can be presented in the following
form
\bea
{\cal L}=-J^\mu(\A'_\mu+a_\mu)-\frac{\kappa}{2}
\varepsilon^{\mu\nu\lambda}a_\mu\p_\nu a_\lambda
+\frac{\ri}{3}\varepsilon^{\mu\nu\lambda}(a_\mu\circ a_\nu\circ
a_\lambda)_{\theta^\star} \label{eq:LAG}
\eea
Here $J^\mu(x)$ is a 3-current consisting of the density $\rho$
and 2-current $j^i$.

Lagrangian (\ref{eq:LAG}) is equivalent to the one given in Ref. 3,
but is based on the use of mechanical variables and as it is claimed
in Ref. 3, this Lagrangian can be considered as an approximation to
the NCCS Lagrangian.

Transition to NCCS theory can be performed simply substituting
the deformed product (\ref{eq:circ}) by the Moyal-Weyl star product
\begin{eqnarray}
(f(x)\star g(x))_\tau&=&\mathrm{e}^{-\mathrm{i}\frac{\tau}{2}
\epsilon^{ik}\partial_{\xi^i}\partial_{\eta^k}}f(x+\xi)\cdot
g(x+\eta)|_{\xi=\eta=0}\nonumber \\
&=&(f(x)\circ g(x))_\tau+ \mathcal{O}(\tau^2)
\end{eqnarray}

This step can be accomplished, taking into account that the LLL
condition forces the area $\Omega$ to be a part of a noncommutative
$x-y$ plane. The  expression
\begin{eqnarray}
\frac{\ri}{\theta}\int_D{\rm d}^2x\epsilon_{ik}(x^i\star x^k)_\theta
=\Omega_\theta
\end{eqnarray}
can be used for an heuristic definition of the area and its
transformation under the map (\ref{eq:map}) accompanied by
the change of the non-commutativity parameter
$\theta\rightarrow\theta^\star=\nu^{-1}\theta$:
\begin{eqnarray}
\Omega_{\theta}\rightarrow\Omega^{\prime}_{\theta^\star}
=\frac{\ri}{ \theta^\star}\int_D{\rd}^2x \epsilon_{ik}
(F^i(\bx)\star F^k(\bx))_{\theta^\star}
\label{eq:mod}
\end{eqnarray}

All the consideration given above can be repeated with the minor
modification of the product:
$(f\circ g)_{\theta^\star}\rightarrow(f\star g)_{\theta^\star}$.
For example for the "covariant curl" one must take
\bea
D^\star_ia_k=\p_ia_k-\ri(a_i\star a_k)_{\theta^\star}
\eea
instead of (\ref{eq:curl}).

Proceeding in this way we arrive at the NCCS Lagrangian
\bea
{\cal L}_{NCCS}=-J^\mu(\A'_\mu+a_\mu)-\frac{\kappa}{2}
\varepsilon^{\mu\nu\lambda}a_\mu\star\bigg\{\p_\nu a_\lambda
-\ri\frac{2}{3}a_\nu\star a_\lambda\bigg\}.
\label{eq:lnc}
\eea
In this expression and hereafter we use the star product with
parameter $\theta^\star$ ($f\star g\equiv(f\star g)_{\theta^\star})$.

\vspace*{5mm}
\noindent
{\bf 3. Chern-Simons Theory in LLL}
\vspace*{3mm}

Up to now we have considered area preserving geometric
transformations, which are supposed to satisfy the constraint
equation
\bea
\rho({{{\bf x}}})+\kappa\ei(\p_i a_k-\ri a_i\star a_k)=0.
\label{eq:GL}
\eea
This constraint have been related to the field theory Lagrangian
(\ref{eq:lnc}). Up to the surface terms this Lagrangian density
is equivalent to
\bea
\mathcal{L}=-J^{i}(A^{\star}_{i}+a_{i})+a_{0}\{J_{0}-\kappa
(\epsilon^{ik}\partial _{i}a_{k}-{\mathrm{i}}\epsilon^{ik}a_{i}
\star a_{k})\}+\frac{\kappa }{2}\epsilon ^{ik}a_{i}\dot{a}_{k}.
\label{eq:Lagrangian2}
\end{eqnarray}

For the interior points of the area $\Omega$ the Euler-Lagrange
equation for the gauge field $a_{\mu}$ reads
\bea
J^{\mu}=-\kappa\{\varepsilon^{\mu\nu\lambda}\partial_{\nu}a_{\lambda}
-\mathrm{i}\varepsilon^{\mu\nu\lambda}a_{\nu}\star a_{\lambda}\}
\eea
and in particular leads to the constraint
\begin{eqnarray}
\Pi_{0}=-J_{0}+\kappa\{\epsilon^{ik}\partial_{i}a_{k}-\mathrm{i}
\epsilon^{ik}a_{i}\star a_{k}\}\approx 0.  \label{eq:gl1}
\end{eqnarray}

The 3-current is not conserved in the usual sense as well as
\begin{eqnarray}
\partial_{\mu}J^{\mu}=\mathrm{i}\kappa\varepsilon^{\mu\nu\lambda}
\partial_\mu \big(a_{\nu}\star a_{\lambda}\big). \label{eq:curdiv}
\end{eqnarray}
Remind, that the 3-current is given by
\begin{eqnarray}
J^0(x)=\rho({{{\bf x}}},t)\hspace*{10mm} J^i(x)={\sum_{\alpha=1}^N}
\dot x^i_\a\delta({{\bf x}}-{{\bf x}}_\a(t)).
\end{eqnarray}

The field $a_{0}$ is Lagrange multiplier providing constraint
(\ref{eq:gl1}) and one can set $a_{0}=0$. In the complex coordinates
Lagrangian reads
\begin{eqnarray}
L&=&\int_{D}\mathrm{d}{{\bf x}}\mathcal{L}=-{\sum_{a=1}^{N}}
\dot{z}_{a}(t)\{{A^\star}_{z}({\bf x}_{a})+a_{z}({\bf x}_{a})\}
-{\sum_{a=1}^{N}}\dot{{\bar{z}}}_{a }(t)\{{A^\star}_{{\bar{z}}}
({{\bf x}}_{a})+a_{{\bar{z}}}({{\bf x}}_{\alpha})\}\nonumber\\
&&+2\mathrm{i}\kappa\int_D\mathrm{d}{{\bf x}}a_{{\bar{z}}}({{\bf x}})
\dot{a}_{z}({{\bf x}})  \label{eq:clag}
\end{eqnarray}
and the Gauss law (\ref{eq:gl1}) is given by
\bea
\rho({\bf x},t)+2\ri\kappa(\p_\bz a_z-\p_z a_\bz)
-4\kappa a_z\star a_\bz=0.\label{eq:glc}
\eea

Lagrangian (\ref{eq:clag}) is a first order in particle velocities
and generates the system of second-class constraints
\begin{eqnarray}
\Pi _{z}^{\alpha}=p_{z}^{\alpha}+A^\star_{z}({{\bf x}}_{\alpha})
+a_z({{\bf x}}_a)\approx 0,  \label{eq:c2}
\end{eqnarray}
\begin{eqnarray}
\Pi_{{\bar{z}}}^{\alpha}=p_{{\bar{z}}}^{\alpha}
+A^\star_{{\bar{z}}}({{\bf x}}_{\alpha})+a_{{\bar{z}}}
({{\bf x}}_{\alpha})\approx 0.  \label{eq:c3}
\end{eqnarray}

Chern-Simons field $a_{\mu}(x)$ must be quantized. The equal time
canonical commutation relation reads
\bea
\lbrack\hat{a}_{z}({{\bf x}}),\hat{a}_{{\bar z}}({{\bf x}}^{\prime })]
=\frac{1}{2\kappa}\delta({{\bf x}}-{{\bf x}}^{\prime}).
\eea
Choosing the holomorphic polarization \cite{nair} we set
\bea
\hat{a}_{\bz}({{\bf x}})=\frac{1}{2\kappa }\frac{\delta }{\delta a_{{z}}
({{\bf x}})}.
\eea

The quantum state vector must satisfy the Gauss law (\ref{eq:glc})
\begin{eqnarray}
\hat\Pi_{0}({{\bf x}})\Phi[a_z;\bx_1,...\bx_N]=0 \label{eq:c1}
\eea
where
\bea
\hat\Pi _{0}({{\bf x}})=\hat J^0({{{\bf x}}})
+2\mathrm{i}\kappa\bigg\{\partial _{{\bar{z}}}a_{z}({{\bf x}})
-\frac{1}{2\kappa}\partial_z\frac{\d}{\d a_z({{{\bf x}}})}\bigg\}
-2a_z({\bf x})\star\frac{\d}{\d a_z({\bf x})}. \label{eq:pi0}
\end{eqnarray}

Together with (\ref{eq:c1}) the state vector is subjected to
the constraint (\ref{eq:c3})
\begin{eqnarray}
\hat\Pi _{{\bar{z}}}^{\alpha}\Phi=\bigg\{\hat p_{{\bar{z}}}^{\alpha}
+\A'_{{\bar{z}}}({{\bf x}}_{\alpha})+\frac{1}{2\kappa}
\frac{\d}{\d a_z({{{\bf x}}}_\a)}\bigg\}\Phi=0.  \label{eq:c3a}
\end{eqnarray}

Equation (\ref{eq:curdiv}) gives
\bea
\bigg\{\p_\mu\hat J^\mu(x)
-\p_t\bigg[a_z(x)\star\frac{\d}{\d a_z(x)}\bigg]\bigg\}\Phi=0. \label{eq:cc1}
\eea

Consider a simplest case of constant ($a_z$-independent) functionals
\bea
\frac{\d}{\d a_z(x)}\Phi_0=0
\eea
Remark, that  in the subspace of constant wave functionals current
is conserved
\bea
\p_\mu\hat J^\mu(x)\Phi_0=0 \label{eq:cc2}
\eea
and one can set
$\hat\rho(x)\equiv\hat J^0(x)=\sma\d({{{\bf x}}}-{\bf x}_\a)$.
Now the Gauss law is reduced to the equation
\bea
\bigg\{\sma \d({\bf x}-{{\bf x}}_a)+2\ri\kappa\p_\bz a_z\bigg\}\Phi_0=0
\eea

Solving the last equation one gets
\bea
a_z({\bf x})=\frac{\ri}{2\pi\kappa}\sma\frac{1}{z-z_\a}.\label{eq:kohno}
\eea
This is a complex connection \cite{kohno} used in the holomorphic gauge
quantization of the non-Abelian CS fields (see {\it e.g.} Ref. 24).
Remark, that the use of this non-Hermitian connection requires
introduction of compensating measure in scalar
products \cite{verlinde,flohr}.

A comment is in order here. It regards the status of the variables $z$
and $\bar z$ in the transformation (\ref{eq:tr1}). The proper approach
is to handle coordinates $z$ and $\bar z$ as independent and impose
the reality condition $z^\star=\bar z$  at the end of calculations.

The wave function $\Phi_0$ depends on the particle coordinates, and
this dependence can be read out from the LLL condition (\ref{eq:c3a}):
\bea
\{\hat p_\bz^\a+\A'_\bz({{{\bf x}}}_\a)\}\Phi_0=0. \label{eq:lll}
\eea
The corresponding solution looks as follows
\bea
\Phi_0({{{\bf x}}}_1,...,{{{\bf x}}}_N)\sim F(z_1,...,z_N)
\re^{-\frac{B^\star}{4}\sma|z_\a|^2}\label{eq:phi}
\eea
with a holomorphic function $F$.

\vspace*{5mm}
\noindent
{\bf 4. Laughlin Wave Function}
\vspace*{3mm}

Wave function of the system of electrons satisfies LLL condition
(\ref{eq:LLL}). The wave function (\ref{eq:phi}) also belongs to
the LLL, but  with respect to the reduced magnetic field
$B^\star=\nu B$. One of the principal assertions of CF approach
is that function $F$ in (\ref{eq:phi}) gives the holomorphic part
of the total LLL wave function (\ref{eq:WF})
\bea
\Psi({\bf x}_1,...,{\bf x}_N)=\re^{-2p\frac{B^\star}{4}\sma|z_\a|^2}
\Phi_{CF}.
\eea

For the electrons in the LLL the filling factor can be defined by
the ratio
\bea
\nu=\frac{N(N-1)}{2J}=\frac{J^\star}{J}.\label{eq:ff2}
\eea
where $J$ is a total angular momentum of the system of electrons.
Strictly speaking definition (\ref{eq:ff2}) is valid in the
thermodynamical limit, but we suppose its validity in our case.

Write down the operator equation
\bea
\hat J=(2p+1)\hat J^\star,
\eea
which corresponds to the fact, that the CF in the magnetic field
$\B'$ occupies the site, which is $(2p+1)$ times larger than Landau
site for the original electron in the magnetic field $B$.

Angular momentum operator for electrons in LLL  is given by
\bea
\hat J=\frac{2}{B}\sum[\hat p^\a_\bz-A_\bz({{{\bf x}}}_\a)]
[\hat p^\a_z-A_z({{{{\bf x}}}}_a)],
\eea
and classically on the constrained manifold
\bea
J(\pi_i^a=0)=\frac{B}{2}\sma|z_\a|^2.
\eea
One easily verifies that analogous expression for the composite
particles is given by
\bea
J^\star(\Pi_i^a=0)&=&\frac{2}{\B'}\sma[p_\bz-A^\star_\bz
({\bf x}_\a)+a_\bz({\bf x}_\a)][p_z-A^\star_z({\bf x}_\a)
+a_z({\bf x}_\a)]\nn\\
&=&\frac{B^\star}{2}\sma|z_\a|^2.
\eea

Using this observation and Eq.(\ref{eq:kohno}) we write down the angular
momentum for the system of composite particles
\bea
\hat J^\star=\frac{2}{\B'}\sma\bigg\{\hat p_\bz-A^\star_\bz({\bf x}_\a)\bigg\}
\bigg\{\hat p_z-A^\star_z({\bf x}_\a)+2\ri p\smb\frac{1}{z_\a-z_\b}\bigg\}.
\eea

The sought for LLL composite fermion wave function is the angular
momentum eigenstate
\bea
\hat J^\star\Phi_{CF}=\frac{N(N-1)}{2}\Phi_{CF},
\eea
and satisfies the Knizhnik-Zamolodchikov \cite{knizhnik} equation
\bea
\bigg\{\hat p_z^\a-A_z^\star+2\ri p\smb \frac{1}{z_\a-z_\b}\bigg\}\Phi_{CF}=
-\ri\smb \frac{1}{z_\a-z_\b}\Phi_{CF}.
\eea
The corresponding solution is given by
\bea
\Phi_{CF}=\prod_{\a<\b}(z_\a-z_\b)^{2p+1}\re^{-\frac{B^\star}{4}
\sma|z_\a|^2}\label{eq:CF}
\eea
yielding the final result -- Laughlin wave function
\bea
\Psi=\prod_{\a<\b}(z_\a-z_\b)^{2p+1}\re^{-\frac{B}{4}\sma|z_\a|^2}
\eea

\vspace*{5mm}
\noindent
{\bf 5. Conclusions}
\vspace*{3mm}

In the present paper we have considered a system of electrons in
the lowest Landau level with the aim to obtain the non-commutative
version of the CS description of quantum Hall effect in terms of
particle variables. We have introduced the area preserving singular
geometric transformations and conclude that the area preservation
condition when interpreted in terms of the QHE filling factor yields
the Gauss law in CS theory with APD as a gauge group. Geometric
transformations are generated by the gauge fields and the Gauss law
is invariant with respect to generalized gauge transformations.
In that part we reproduce corresponding conclusions given in Ref. 3.

As a further step we have proposed the modification of the area
transformation rule. This modification is presented as an
heuristic tool without any special justification. Despite of that,
this {\it Ansatz} leads to the Gauss law in the form adopted in
the NCCS theory and permits to write down corresponding Lagrangian.
This Lagrangian describes particles in the effective (reduced)
magnetic field and interacting with the non-commutative CS field.
Developed scheme corresponds to the composite fermion picture in
the non-commutative CS theory.

The quantum state vector depends on the gauge field configurations
and particle coordinates. Dynamics is governed by the system of
constraints imposed on the state vector. These conditions are Gauss
law and constraints expressing vanishing of kinetic momenta of
particles in the LLL. We have examined the self-consistent solution
corresponding to the constant (gauge field independent) wave
functional. In that case one may identify the particle density as a
time component of the conserved local 3-current and consideration
is reduced to the linear CS theory in the holomorphic gauge.

The detailed form of the wave function was determined considering
the total angular momentum of the system of composite fermions.
This permitted to express the  corresponding solution in the form
of the Laughlin wave function.

The proposed scheme seems to be equivalent to the hydrodynamical
formulation, but the use of mechanical variables permits to
reconstruct the structure of the many-electron wave function in
the composite fermion approach.

\vspace*{5mm}
\noindent
{\bf Acknowledgements}
\vspace*{3mm}

Authors thank P. Sorba and A. Tavkhelidze for the interest and
encouraging remarks. M.E. is  grateful to P. Sorba for his hospitality
at LAPTH (Annecy), where the part of the present work was done.
Work was supported in part by the grant INTAS-GEORGIA 97-1340 and by
SCOPES under grant 7GEPJ62379.


\begin{thebibliography}{99}
\bibitem{ezawa}
Z. F. Ezawa, {\it Quantum Hall Effects: Field Theoretical Approach
and Related Topics}, (World Scientific, Singapore, 2000).

\bibitem{jackiw}
R. Jackiw, {\it Physical instances of noncommuting coordinates},
hep-th/0110057.

\bibitem{susskind}
L. Susskind, {\it The quantum Hall fluid and non-commutative
Chern-Simons theory}, hep-th/0101029.

\bibitem{hellerman}
S. Hellerman and M. van Raamsdonk, {\it Quantum Hall physics
equals noncommutative field theory}, hep-th/0103179.

\bibitem{polychronakos1}
A. P. Polychronakos, {\it Quantum Hall states as matrix
Chern-Simons theory}, hep-th/0103013.

\bibitem{duval}
C. Duval and P. A. Horvathy, {\it Exotic Galilean symmetry in the
noncommutative plane and quantum Hall effect}, hep-th/0106089.

\bibitem{pasquier}
V. Pasquier, {\it Skyrmions in the quantum Hall effect and
noncommutative solitons}, hep-th/0007176.

\bibitem{polychronakos2}
A. P. Polychronakos, {\it Quantum Hall states on the cylinder as
unitary matrix Chern-Simons theory}, hep-th/0106011.

\bibitem{morariu}
B. Moraiu and A. P. Polychronakos, {\it Finite noncommutative
Chern-Simons with a Wilson line and the quantum Hall effect},
hep-th/0106072.

\bibitem{guralnik}
Z. Guralnik, R. Jackiw, S.-Y. Pi and A. P. Polychronakos,
{\it Testing non-commutative QED, constructing non-commutative MHD},
hep-th/0106044.

\bibitem{laughlin} R. B. Laughlin, {\it Phys. Rev. Lett.} {\bf50},
1395 (1983).

\bibitem{bahcall}
S. Bahcall and L. Susskind, {\it Int. J. Mod. Phys.}
{\bf B5}, 2735 (1991).

\bibitem{jain}
J. K. Jain, {\it Phys. Rev. Lett.} {\bf 63}, 199 (1989).

\bibitem{lopez}
A. Lopez and E. Fradkin, {\it Phys. Rev.} {\bf B44}, 5246 (1991).

\bibitem{zhang}
S. C. Zhang, {\it Int. J. Mod. Phys.} {\bf B6}, 25 (1992).

\bibitem{manvelyan}
R. Manvelyan and R. Mkrtchyan, {\it Phys. Lett.} {\bf B327}, 47 (1994).

\bibitem{eliashvili00}
M. Eliashvili and G. Tsitsishvili, {\it Int. J. Mod. Phys.}
{\bf B14}, 1429 (2000).

\bibitem{dirac}
P. Dirac, {\it Lectures on Quantum Mechanics}
(Belfer Graduate School of Science, Yeshiva University, New York, 1964),

\bibitem{dunne}
G. V. Dunne, R. Jackiw and C. A. Trugenberger,
{\it Phys. Rev.} {\bf D41}, 661 (1990).

\bibitem{cappelli}
A. Cappelli, C. Trugenberger and G. Zemba, {\it Nucl. Phys.}
{\bf B396}, 465 (1993).

\bibitem{karabali}
D. Karabali, {\it Nucl. Phys.} {\bf B419}, 437 (1994).

\bibitem{nair}
M. Bos and V. P. Nair, {\it Int. J. Mod. Phys.}
{\bf A5}, 959 (1990).

\bibitem{kohno}
T. Kohno, {\it Ann. Inst. Fourrier} (Grenoble) {\bf 37.4}, 139 (1987).

\bibitem{lee}
T. Lee and P. Oh, {\it Ann. Phys.} (N.Y) {\bf 235}, 413 (1994).

\bibitem{verlinde}
E. Verlinde, {\it A note on braid statistics and the non-abelian
Aharonov-Bohm effect}, in {\it Modern Quantum Field Theory}
ed. A. Das {\it at al.} (World Scientific, Singapore 1991)

\bibitem{flohr}
M. Flohr and R. Varnhagen, {\it Journ. of Phys.} {\bf A27}, 3999 (1994).

\bibitem{knizhnik}
V. Knizhnik and A. Zamolodchikov, {\it Nucl. Phys.}
{\bf B247}, 139 (1984)
\end{thebibliography}
\end{document}